# Intrinsic defect engineering of CVD grown monolayer MoS$_2$ for tuneable functional nanodevices

*Irfan H. Abidi,*[1,$,*] *Sindhu Priya Giridhar,*[1,$] *Jonathan O. Tollerud,*[2] *Jake Limb,*[3] *Aishani Mazumder,*[1] *Edwin LH Mayes,*[4] *Billy J. Murdoch,*[4] *Chenglong Xu,*[5] *Ankit Bhoriya,*[1] *Abhishek Ranjan,*[6] *Taimur Ahmed,*[1,7] *Yongxiang Li,*[1] *Jeffrey A. Davis,*[2] *Cameron L. Bentley,*[3] *Salvy P. Russo,*[8] *Enrico Della Gaspera,*[8] *Sumeet Walia*[1,*]

[1] School of Engineering, RMIT University, 124 La Trobe Street, Melbourne 3000, Australia

[2] Optical Sciences Centre, Swinburne University of Technology, Victoria 3122, Australia

[3] School of Chemistry, Monash University, Clayton 3800, Victoria, Australia

[4] RMIT Microscopy and Microanalysis Facility, RMIT University, Melbourne 3000, Australia

[5] Micro Nano Research Facility, RMIT University, Melbourne 3000, Australia

[6] School of Engineering, Monash University Malaysia, Subang Jaya 47500, Malaysia

[7] School of Computing Sciences, Pak-Austria Fachhochschule Institute of Applied Sciences and Technology, Haripur, 22620, Pakistan

[8] School of Science, RMIT University, 124 La Trobe Street, Melbourne 3000, Australia

[*] *Correspondence to: Irfan H. Abidi (*Irfan.haider.abidi@rmit.edu.au*) and*

*Sumeet Walia (*sumeet.walia@rmit.edu.au)

[$] *Equal contributing authors.*

## Abstract

Defects in atomically thin materials can drive new functionalities and expand applications to multifunctional systems that are monolithically integrated. An ability to control formation of defects during the synthesis process is an important capability to create practical deployment opportunities. Molybdenum disulfide (MoS$_2$), a two-dimensional (2D) semiconducting material harbors intrinsic defects that can be harnessed to achieve tuneable electronic, optoelectronic, and electrochemical devices. However, achieving precise control over defect formation within monolayer MoS$_2$, while maintaining the structural integrity of the crystals remains a notable challenge. Here, we present a one-step, in-situ defect engineering approach for monolayer MoS₂ using a pressure dependent chemical vapour deposition (CVD) process. Monolayer MoS₂ grown in low-pressure CVD conditions (LP-MoS₂) produces sulfur vacancy ($V_s$) induced defect rich crystals primarily attributed to the kinetics of the growth conditions. Conversely, atmospheric pressure CVD grown MoS₂ (AP-MoS₂) passivates these $V_s$ defects with oxygen. This disparity in defect profiles profoundly impacts crucial functional properties and device performance. AP-MoS₂ shows a drastically enhanced photoluminescence, which is significantly quenched in LP-MoS₂ attributed to in-gap electron donor states induced by the $V_s$ defects. However, the n-doping induced by the $V_s$ defects in LP-MoS₂ generates enhanced photoresponsivity and detectivity in our fabricated photodetectors compared to the AP-MoS₂ based devices. Defect-rich LP-MoS₂ outperforms AP-MoS₂ as channel layers of field-effect transistors (FETs), as well as electrocatalytic material for hydrogen evolution reaction (HER). This work presents a single-step CVD approach for in-situ defect engineering in monolayer MoS₂ and presents a pathway to control defects in other monolayer material systems.

**Introduction**

The next-generation of multifunctional, monolithic semiconducting technology integration relies on the ability to readily tune the properties of 2D semiconductors.[1-4] Customizing the electronic and optical properties of semiconductor crystals through defect engineering is a well-established strategy in the semiconductor industry.[5, 6] While recently a large volume of research continues into 2D materials and their intriguing properties, a crucial step to make these materials practically viable is an ability to control the formation of defects and therefore their intrinsic properties.[7-13] Among the many 2D materials being studied, transition metal dichalcogenides (TMDs), such as monolayer MoS₂ have shown great potential as an alternative semiconducting material to existing silicon-based electronics and optical technologies, owing to their unique characteristics. In fact, MoS₂ has a direct and tuneable bandgap (~1.8-1.9 eV), moderate carrier mobility, atomic level thickness (~ 0.7 nm) and is mechanically stable.[13-16]

Such properties make it an attractive proposition for next generation miniaturized functional devices such as photodetectors,[12, 17] inverters,[18] field-effect transistors (FETs),[19, 20] sensors,[21] electrocatalysts[22, 23] and memory storage devices.[24, 25] An intriguing aspect of such TMDs is their performance reliance on the quality of the 2D atomic crystals, which can be influenced by atomic defects such as sulfur vacancies ($V_s$) and atomic substitutions.[9, 12] Nonetheless, the $V_s$ defects are the most prevalent intrinsic structural defects that tend to manifest in $MoS_2$ crystals, primarily due to the formation of $V_s$ is considerably energetically more favourable compared to other structural defects.[9, 26] Therefore, implementing a viable defect engineering approach to control these intrinsic $V_s$ defects in $MoS_2$ can enable harnessing the functionality of these 2D materials for enhanced device performance.

Achieving reliable and scalable defect engineering in monolayer $MoS_2$ poses significant challenges. Various chemical and physical approaches have been reported to manipulate defects within the 2D lattice including electron/ion beam and UV photon irradiation,[27-29] laser processing,[30] plasma treatment,[31, 32] chemical functionalization[18, 33] and thermal annealing in hydrogen.[34, 35] All of these techniques are based on a two-step process, requiring post-synthesis processing and often associated with detrimental effects on monolayer $MoS_2$ due to the usually aggressive nature of the techniques. This often results in structural degradation, uncontrolled surface modification, contamination and doping leading to inhomogeneous and unpredictable properties.[29] Furthermore, an additional post-processing step results in increased cost, energy use and reduced compatibility with complementary metal–oxide–semiconductor (CMOS) processes and variations in the performance, which subsequently impedes their scope of implementation into next-generation devices.[36] Thus, alternative cost-effective, one-step strategy to enable precise defect control while maintaining scalability and preserving structural integrity is imperative to extract end-user benefit of the unique properties on offer from $MoS_2$.

In this work, we demonstrate a simple, effective and scalable strategy to inherently regulate the formation and/or the passivation of $V_s$ defects on demand within monolayer $MoS_2$, by tuning the pressure during CVD growth. Our investigation reveals that $MoS_2$ monolayers grown by low-pressure CVD (LP-$MoS_2$) exhibit a higher density of $V_s$ defects, due to the fast kinetics achieved during the growth process. While $MoS_2$ monolayers grown using atmospheric pressure CVD (AP-$MoS_2$) contain fewer $V_s$ defects, owing to oxygen passivation driven by the residual oxygen present in ambient conditions. We show that these controlled and deliberate defect variations can be used to modulate optical, electronic and electrocatalytic properties of as-grown $MoS_2$ monolayers. AP-$MoS_2$ demonstrates two orders of magnitude higher photoluminescence (PL) as compared to LP-$MoS_2$, originating from the passivation of $V_s$ defects which act as photogenerated carrier traps. To further reveal the significant differences in key properties, we fabricated a series of $MoS_2$ based devices (photodetectors, FETs, and water splitting). LP-$MoS_2$ based photodetectors exhibits one order of magnitude higher photoresponsivity and detectivity compared to AP-$MoS_2$ devices, which correlate to the n-doping induced by $V_s$ defects. Similarly, higher carrier density and a lower contact resistance induced by $V_s$ defects accumulatively contributed to an enhanced on-state current, and superior electron mobility in LP-$MoS_2$ FETs. Moreover, the defect rich basal plane of LP-$MoS_2$ leads to an enhanced catalytic activity for hydrogen evolution reaction (HER) across the whole surface, and not just on the edges. Therefore, our work provides a simple, one-step CVD approach to readily control the intrinsic atomic defects in monolayer $MoS_2$, leading to a reliable modulation of critical electronic, optoelectronic and catalytic characteristics that can be easily adapted based on specific applications. This strategy can be adapted to enable precise defect engineered monolayer TMDs, with customizable properties for targeted applications in atomically thin optoelectronics, electronics neuromorphic, and catalytic devices. It also opens new avenues for monolithic integration of multifunctional devices based on monolayer $MoS_2$.

**Results and Discussion**

Figure 1a shows a schematic of the two-temperature-zone CVD system utilised for the synthesis of MoS$_2$ monolayers in this study under two different growth configurations (see methods section for further details). AP-MoS$_2$ is grown using atmospheric- pressure conditions (760 torr), and LP-MoS$_2$ growth is performed under relatively low-pressure conditions (1 Torr), maintained using a vacuum pump. For both the growth conditions, the sulfur (S) powder is vaporised at 180 °C in zone-1 and the vapours are carried downstream by a non-reactive carrier gas to the growth zone-2 where Mo-precursor precoated 300 nm SiO$_2$/Si substrates are placed at 750 °C (see methods section for details). Figure 1 b-c and S1 (supporting information) show the optical images of the as-grown single crystal triangular domains of AP-MoS$_2$ and LP-MoS$_2$, respectively. It is revealed that AP-MoS$_2$ domains are mostly of regular equilateral triangular shape with smooth edges, while LP-MoS$_2$ domains show triangular shape with slightly concave and jagged edges. The discrepancy in shape of the MoS$_2$ indicates the difference in the growth kinetics under two different growth conditions. The dendritic edges of as-grown LP-MoS$_2$ 2D crystals indicate the instability during crystal growth arising from the fast kinetics achieved during low pressure CVD conditions.[37, 38] This is in contrast with the atmospheric CVD conditions, where the growth follows a thermodynamically stable route resulting in compact and smooth edges of the 2D crystals.[39] Despite the difference in crystal shapes, the atomic force microscopy (AFM) analysis shown in Figure 1 d-e reveals a similar thickness of ~ 0.7-0.8 nm for the MoS$_2$ grown using both conditions, indicating no notable thickness variations and is consistent with the reported thickness of monolayer MoS$_2$ films.[38, 40] To further reveal the atomic structure of the as-grown MoS$_2$ domains, high resolution transmission electron microscopy (HRTEM) images are shown in Figure 1 f-g. These images confirm the typical honeycomb lattice arrangement of MoS$_2$ (100) with, interplanar spacings of 0.27 nm for both AP-MoS$_2$ and LP-MoS$_2$. Moreover, selected-area electron diffraction (SAED) analysis taken for both the MoS$_2$ show sixfold-symmetric diffraction points, which corresponds to the single crystal hexagonal lattice structure of monolayer MoS$_2$,[40] indicating the high crystallinity of the as-grown crystals. The presence of defects is revealed using the atomic resolution images shown in Figure 1g indicates AP-MoS$_2$ as a largely defect-free lattice and LP-MoS$_2$ with the presence of few atomic vacancy defects as indicated by arrows.

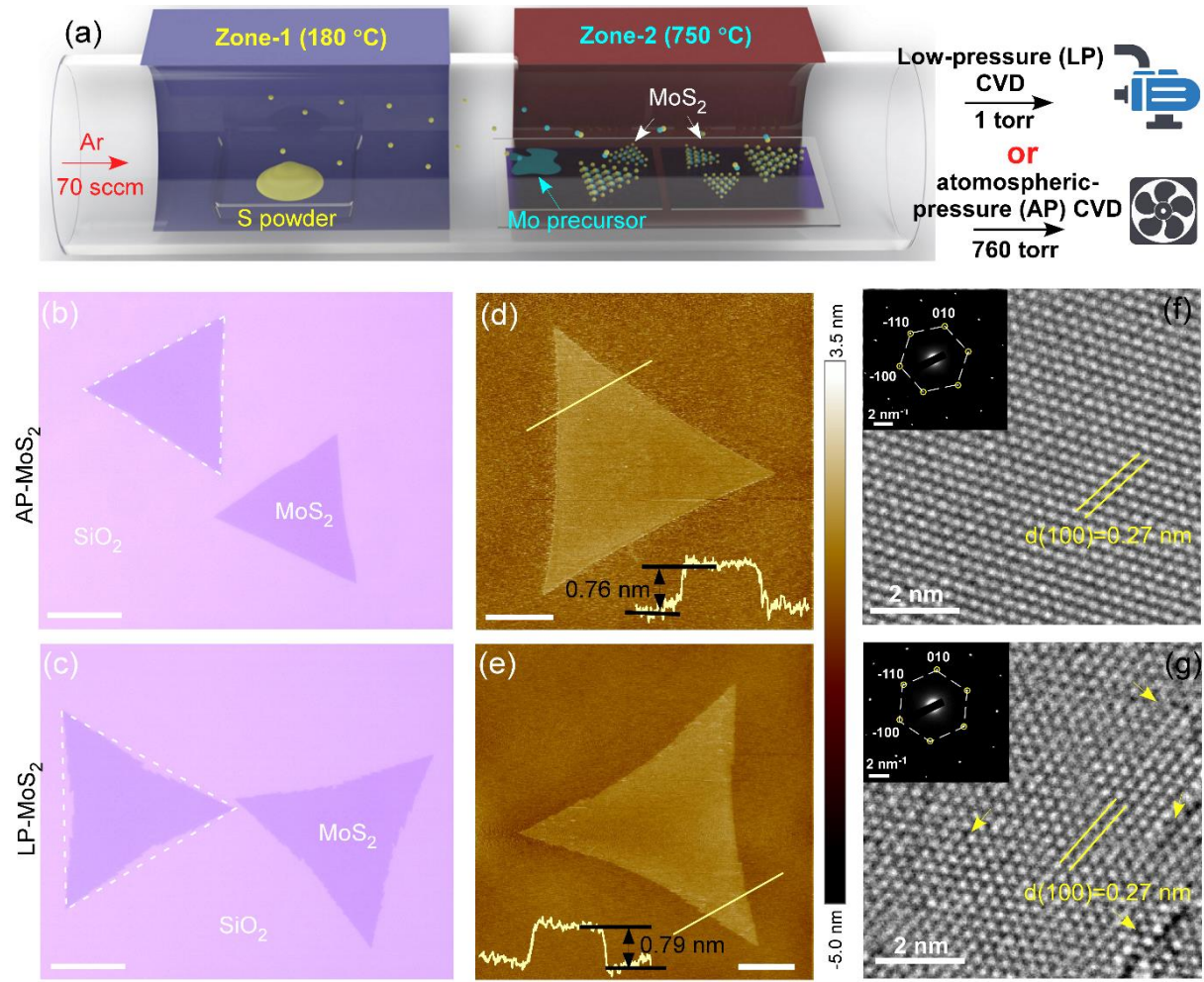

**Figure 1.** (a) Schematic of the CVD processes adopted in this study. (b-c) Optical images of typical single crystals of as-grown AP-MoS$_2$ and LP-MoS$_2$, respectively. The scale bar is 20 μm. (d-e) AFM images of a representative AP-MoS$_2$ and LP-MoS$_2$ crystal, respectively. Both crystals show the thickness of a typical monolayer MoS$_2$ (~ 0.7-0.8 nm). The scale bar is 10 μm. (f-g) High resolution TEM images of AP-MoS$_2$ and LP-MoS$_2$ crystal, respectively. The scale bar is 2 nm. The arrows indicate some of the atomic defects in the LP-MoS$_2$ crystal. The insets show the corresponding SAED pattern.

To further probe the quality and compositional/chemical characteristics of the as-grown single-crystal $MoS_2$ domains, we obtained the Raman spectra of the films, shown in Figure 2a. Raman spectral analysis is a powerful technique used to investigate structural inhomogeneities such as atomic defects, doping or strain in 2D lattices.[41] Raman spectra obtained from both types of $MoS_2$ show typical $E^1_{2g}$ and $A_{1g}$ bands associated with in-plane and out of plane vibrations.[41] For AP-$MoS_2$, $E^1_{2g}$ and $A_{1g}$ peaks appear at ~384 and ~404 cm$^{-1}$, respectively, with a frequency difference (Δ) of ~20 cm$^{-1}$, which is consistent with the values reported for CVD grown monolayer $MoS_2$ flakes.[39, 40] However, LP-$MoS_2$ exhibits a redshift and slight broadening in the peaks associated with both $A_{1g}$ (~403 cm$^{-1}$) and $E^1_{2g}$ (~381 cm$^{-1}$) vibrational modes, as compared to that of AP-$MoS_2$. While the AFM analysis confirmed the monolayer thickness of LP-$MoS_2$, the Δ of ~22 cm$^{-1}$ can be attributed to the formation of structural defects within $MoS_2$ lattice.[28, 31, 42] This is consistent with the atomic defects observed in HRTEM image of LP-$MoS_2$, shown in Figure 1g. The redshift in Raman modes suggests the formation of $V_s$ defects within $MoS_2$ lattice, since the presence of $V_s$ defects weakens the lateral vibrations within the Mo-S bonds, resulting in weaker restoring force constant (phonon softening), translating it into redshift of the $E^1_{2g}$ phonon frequency.[41] Fitting of the Raman peaks further reveals the presence of defects related to Raman modes such as longitudinal optical branch (LO (M) at ~377 cm$^{-1}$ and out-of-plane optical branch (ZO (M) at ~409 cm$^{-1}$, respectively, in LP-$MoS_2$, as shown in Figure 2b. An intense peak of LO(M) defect mode, which is primarily associated with $V_s$,[31, 34] further confirms the presence of sulfur vacancies in LP-$MoS_2$. In contrast, the AP-$MoS_2$ exhibits negligible Raman modes related to defects suggesting minimal presence of $V_s$ defects. It is important to note that $V_s$ defects can act as an electron donor sites due to the unsaturated bonds, inducing n-doping effect in the monolayer $MoS_2$.[43] Such n-doping effect is also evident by the slight redshift (404 cm$^{-1}$ to 403 cm$^{-1}$) of $A_{1g}$ band in LP-$MoS_2$ sample relative to AP-$MoS_2$, which can be translated into additional electron concentration of

~ 4.5 (±1) x $10^{12}$ cm$^{-2}$ in LP-MoS$_2$ sample (as 2.22 cm$^{-1}$ Raman shift in A$_{1g}$ band translate into 1x10$^{13}$ cm$^{-2}$ carrier concentration).[42] Therefore, Raman analysis clearly suggest that LP-MoS$_2$ is significantly n-doped due to the presence of V$_s$ defects.

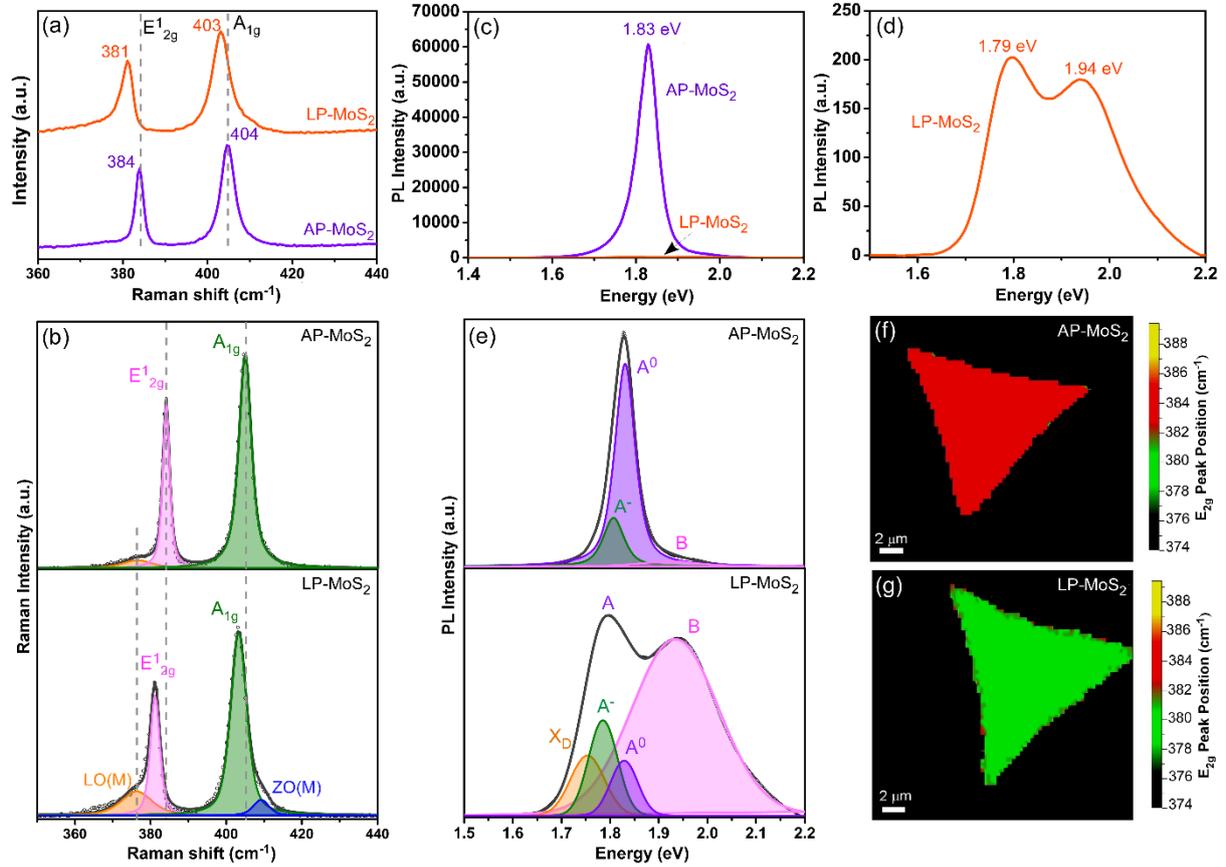

**Figure 2.** Raman and PL analysis of AP-MoS$_2$ and LP-MoS$_2$ samples. (a) Raman spectra obtained from as-grown AP-MoS$_2$ and LP-MoS$_2$ monolayers. The redshift in E$^1_{2g}$ and A$_{1g}$ is guided by the dotted lines. (b) The fitted Raman spectra of AP-MoS$_2$ and LP-MoS$_2$, showing peaks associated with E$^1_{2g}$, A$_{1g}$, LO(M) and ZO(M) Raman modes. (c) PL spectra of as-grown AP-MoS$_2$ and LP-MoS$_2$ monolayers, showing a significant PL enhancement (~300 times) in AP-MoS$_2$. (d) Enlarged view of the PL spectrum of LP-MoS$_2$. (e) The deconvoluted PL spectra of AP-MoS$_2$ and LP-MoS$_2$, showing peaks associated with neutral exciton (A$^0$), negatively charged trion (A$^-$), B-exciton (B), and defect-induced exciton (X$_D$). (f-g) Raman mapping of E$^1_{2g}$ mode for AP-MoS$_2$ and LP-MoS$_2$ single crystal domains.

To further investigate the role of defects within MoS$_2$ crystals, we obtained the room temperature photoluminescence (PL) spectra, and we observed a stark contrast in the PL signal between AP-MoS$_2$ and LP-MoS$_2$. The PL intensity of AP-MoS$_2$ is significantly enhanced by two orders of magnitude, (~300 times) compared to that of LP-MoS$_2$, as evidenced from Figure 2c, where the two spectra are plotted in the same scale. The PL intensity of monolayer MoS$_2$ is proportional to the radiative transition of neutral excitons and is highly sensitive to the structural defects prevalent within the crystal system. Atomic defects such as V$_s$ can create defects states within the bandgap of MoS$_2$, promoting lower energy transitions and even mediating non-radiative recombination that quench the PL intensity.[10, 34] Therefore, the PL reduction in LP-MoS$_2$ further indicates the formation of V$_s$ defects, while the intense PL spectrum of AP-MoS$_2$ implies minimal defect states formation in the crystal, which is consistent with our Raman analysis and XPS analysis (discussed later).

The line shape of the two spectra (Figure 2d) shows clear differences. The AP-MoS$_2$ PL spectrum shows a strong emission peak at ~1.83 eV and a weaker shoulder peak at ~1.98 eV, corresponding to A and B excitonic transitions in monolayer MoS$_2$ originating from direct band gap transition and valence band splitting caused by strong spin–obit coupling, respectively.[34] These peaks are not just reduced in intensity, but also slightly shifted to ~1.79 eV and 1.94 eV, respectively, for LP-MoS$_2$. To investigate the origin of this discrepancy in PL spectra, we fitted the PL peaks into further contributing excitonic peaks to account for the combination of transitions, shown in Figure 2e. The A excitonic peak is the combination of neutral exciton (A$^0$) and a negatively charged trion (A$^-$) which is formed due to the interaction of electron with the neutral exciton.[32] The V$_s$ structural defects in MoS$_2$ act as electron donors due to unsaturated bonds, resulting in n-doping of MoS$_2$ and facilitate the formation of trions.[43, 44] The A excitonic peak in the PL spectrum of AP-MoS$_2$ is fitted into a dominant neutral exciton A$^0$ and a weaker A$^-$ peak, which suggests that due to the absence of defect states, the transitions are mainly dominated by radiative recombination occurring at ground state, and hence emitting a strong $A^0$-peak and relatively weak $A^-$ and B-peaks.[34, 44] In contrast, the PL spectra of LP-MoS$_2$ has been fitted with four excitonic peaks, with an additional peak at ~1.75 eV, assigned as $X_D$, which is associated with defect-induced emission originating from the excitonic transitions occurring from the defect induced in-gap states lying within the band gap of MoS$_2$.[31, 34] Furthermore, the absorbance spectra shown in Figure S3 validate the formation of in-gap states by showing relatively sharper peaks of corresponding exciton-A and exciton-B in AP-MoS$_2$, while the peak broadens for these excitonic transition in LP-MoS$_2$ spectrum. This is also consistent with a defect-induced quenching of PL due to the increase in non-radiative recombination processes. Furthermore, the $A^-$ trion peak is dominant as compared to $A^0$ neutral exciton peak, evident from the ratio of integrated intensity of $A^-$ trion to $A^0$ exciton, $I(A^-)/I(A^0)$, which increased from 0.25 in AP-MoS$_2$ to 1.84 in LP-MoS$_2$. The increased integrated intensity ratio of $A^-$ trion to $A^0$ exciton in LP-MoS$_2$ indicates the n-doping caused by the formation of the $V_s$ structural defects, as reported previously.[45] Figure 2f and S4 present Raman mapping of the $E^1_{2g}$ peak positions, $A_{1g}$ peak width, and the frequency difference ($\Delta$) of $A_{1g}$ and $E^1_{2g}$ modes revealing the microscopic uniformity with consistent shift in the Raman modes across each type of AP-MoS$_2$ and LP-MoS$_2$ crystals.

To correlate the change in electronic structure of MoS$_2$ due to $V_s$ with the excitonic transitions in PL spectra, we measured exciton decay lifetimes in both types of MoS$_2$ using time-resolved photoluminescence (TRPL), shown in Figure 3a. To explore the difference in decay rate, we extract decay constants by fitting a bi-exponential decay function. A fast decay ($\tau_1$) and a slow decay ($\tau_2$) decay constant is observed for both samples, measuring 0.20 ns and 1.14 ns for AP-MoS$_2$ and 0.18 ns and 1.90 ns for LP-MoS$_2$, respectively. The sub-ns fast response ($\tau_1$) can be attributed either to the radiative decay of the direct exciton transition or and trapping in defect states, whereas the slower response ($\tau_2$) originates solely from the trapped excitons that are freed from their traps via thermal energy (under ambient conditions).[46] The high density of V$_s$ defects in LP-MoS$_2$ creates in-gap trap states for the excitons, and hence a larger fraction of the excitons quickly settle in traps prior to the radiative decay. These traps have longer decay lifetimes, and introduce additional non-radiative decay channels, which reduces the TRPL yield due to increased defects states leading to higher non-radiative decay thereby shortening $\tau_1$ in LP-MoS$_2$.[34] The higher concentration of trap states in LP-MoS$_2$ leads to more efficient re-trapping of thermally freed excitons, increasing the $\tau_2$ decay lifetime of the charge carriers due to increased carrier lifetime in trap states In AP-MoS$_2$, the absence of in-gap states reduces the trap sites, leading to less efficient re-trapping, more rapid depletion of trapped excitons, and thus a shorter $\tau_2$.[34]

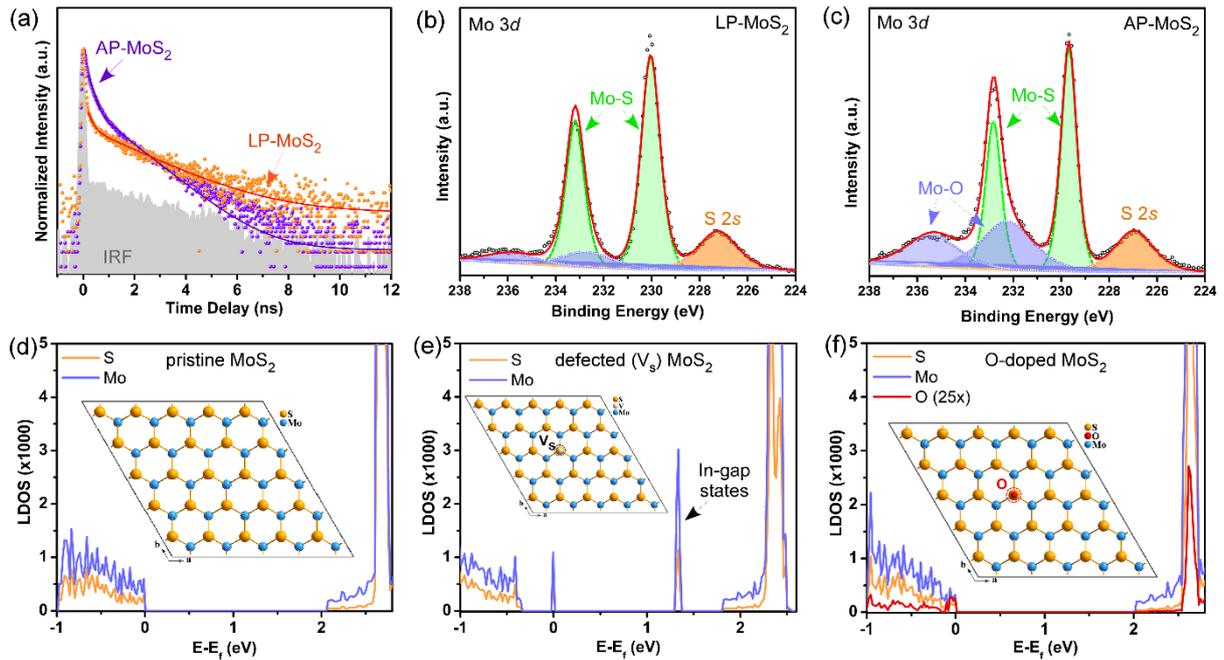

**Figure 3.** (a) Time-resolved PL measurements for AP-MoS$_2$ and LP-MoS$_2$ samples, showing longer decay lifetimes for LP-MoS$_2$ attributed to V$_s$ defects induced trap-states within the bandgap (b-c) XPS spectra of the Mo 3d for LP-MoS$_2$ and AP-MoS$_2$ monolayers, showing characteristic peaks of Mo(IV) 3d doublet (green) and S 2s (orange), associated with MoS$_2$ in both spectra. The evolution of intense peaks related to Mo(VI) 3d doublet (blue) in AP-MoS$_2$, signify the presence of Mo-O bonds in AP-MoS$_2$, suggesting oxygen passivation of V$_s$. (d-f) Plots for the local density of states (LDOS) for pristine MoS$_2$, V$_s$ defect rich MoS$_2$, and MoS$_2$ with oxygen passivated V$_s$. Inset show atomic structural model obtained from DFT, respectively.

To further investigate the chemical state and composition of the AP-MoS$_2$ and LP-MoS$_2$, X-ray photoelectron spectroscopy (XPS) measurements are performed. Figure 3 b,c shows the XPS spectra for both the as-grown monolayer MoS$_2$, and relative fittings in the Mo3d / S2s region. Main peaks are observed at 226.9 eV, 229.6 eV, and 232.7 eV, which correspond to S 2s, Mo 3d doublet of Mo(IV) 3d$_{5/2}$ and Mo(IV) 3d$_{3/2}$, which are consistent with MoS$_2$.[40] Additional peaks at 232.3 and 235.4 eV are also observed which correspond to another Mo 3d doublet (Mo(VI) 3d$_{5/2}$ and Mo(VI) 3d$_{3/2}$), indicating the existence of MoO$_x$ phase.[10, 29] While the weak intensity of these Mo–O peaks is quite weak in LP-MoS$_2$ , the Mo-O contribution becomes stronger in the AP-MoS$_2$. These peaksin LP-MoS$_2$ spectrum can be associated the contribution of physiosorbed oxygen from ambient environment,[10, 43] these peaks are quite intense for AP-MoS$_2$, suggesting the presence ofbut also of Mo-O ionic bond,bonds, which in the AP-MoS$_2$ could presumably originatingoriginate from the oxygen passivation of MoS$_2$ surfaces.[43] Furthermore the extracted S/Mo ratios from XPS analysis is higher in AP-MoS$_2$ (~1.91) as compared to that of LP-MoS$_2$ (1.71) suggesting again the presence of high density of V$_s$ defects in LP-MoS$_2$. These data collectively establish that during low pressure CVD growth conditions, a large amount of V$_s$ is formed and preserved in monolayer MoS$_2$, however, the presence of residual oxygen during the atmospheric pressure CVD growth conditions,[47, 48] promoted an in-situ oxygen passivation mechanism of these V$_s$. Consequently, this structural variation modulates the electronic structure of monolayer MoS$_2$, also validated by work function measurements using Kelvin probe force microscopy (KPFM), (Figure S5, Supporting Information ). The LP-MoS$_2$ exhibits a lower work function of ~ 4.78 eV as compared to the work function measured for AP-MoS$_2$ (~ 4.95 eV). This implies that the formation of V$_s$ defects in LP-MoS$_2$ shifts the Fermi level closer towards the conduction band (due to n-doping), whereas the oxygen passivation in AP-MoS$_2$ results in the Fermi level shift towards the intrinsic level of largely defect free MoS$_2$.[29, 49]

Density functional theory (DFT) calculations were conducted to investigate the role of V$_s$ defects and their oxygen passivation in monolayer MoS$_2$ and further corroborate our experimental findings. The local density of states (LDOS) along with the atomic structure model for pristine MoS$_2$, defected MoS$_2$ with V$_s$, and oxygen passivated MoS$_2$ are shown in Figure 3d-f. LDOS calculation reveal that the introduction of V$_s$ defects in MoS$_2$ lattice increases the electron concentration and induces electron donor states within the bandgap of MoS$_2$ (Figure 3b), so called in-gap states,[11] which are absent in LDOS of pristine MoS$_2$ (Figure 3a). The passivation of such V$_s$ sites with oxygen atoms removes the in-gap donor states, retaining LDOS similar to pristine MoS$_2$, as shown in Figure 3f. The oxygen passivation therefore promotes electron depletion and counterbalance the n-doping effect caused by V$_s$.[50] These theoretical insights are well aligned with our aforementioned experimental interpretation obtained from the Raman/PL, TRPL and work function measurements.

We now demonstrate the role of these defects by fabricating and testing devices based on AP and LP-MoS$_2$ as functionally active elements in photodetectors, FETs and HER. Figure 4a shows a 3D schematic and an optical image (inset) of a typical MoS$_2$ device used for photodetector and FET measurements. As grown monolayer AP-MoS$_2$ and LP-MoS$_2$ growth on 300 nm SiO$_2$/Si substrates with lateral Ni/Au top contacts are utilised. This device structure enables to test MoS$_2$ as a two-terminal photodetector, and a three-terminal FETs using the Si (p$^{++}$) as a back gate. We started our investigation with photodetectors, by measuring the the time-dependant photoresponse of AP-MoS$_2$ and LP-MoS$_2$ intermittent pulsed illumination (time period of light pulse) at a constant readout voltage (V$_{DS}$) of 2V (Figure 4b-c). Under the excitation wavelengths of both 660 nm and 565 nm (with an incident power density of 3 mW/cm$^2$), the LP-MoS$_2$ based devices exhibit approximately two orders of magnitude higher photocurrent ($I_{photo}$) as compared to that of AP-MoS$_2$ devices. As observed from the figures both the variants of MoS$_2$ demonstrate cyclic repeatability of the photodetection. Interestingly, the dark current ($I_{dark}$) of LP-MoS$_2$ is also an order of magnitude higher than AP-MoS$_2$ devices as shown in Figure S6, consistent with their stronger n-type character and higher electrical conductivity. To better compare the two types of MoS$_2$, we calculated the typical figures of merit used for photodetectors: responsivity and detectivity. These are plotted against varying illumination power densities in Figure 3d. The LP-MoS$_2$ device demonstrates at least one order higher magnitude of responsivity and detectivity as compared to those of the AP-MoS$_2$ at all measured power densities. These results can be explained by the n-doping in LP-MoS$_2$, which contributes to the higher carrier density leading to higher $I_{dark}$ and $I_{photo}$.[29, 51] In contrast, the annihilated n-doping effect in AP-MoS$_2$, results in lower carrier concentration and thus inducing lower $I_{dark}$ and $I_{photo}$.[51] Importantly, the slight decreasing trend in photoresponsivity of LP-MoS$_2$ with increasing power density can be attributed to the saturation of the V$_s$ induced trap states present in defect rich LP-MoS$_2$.[17]

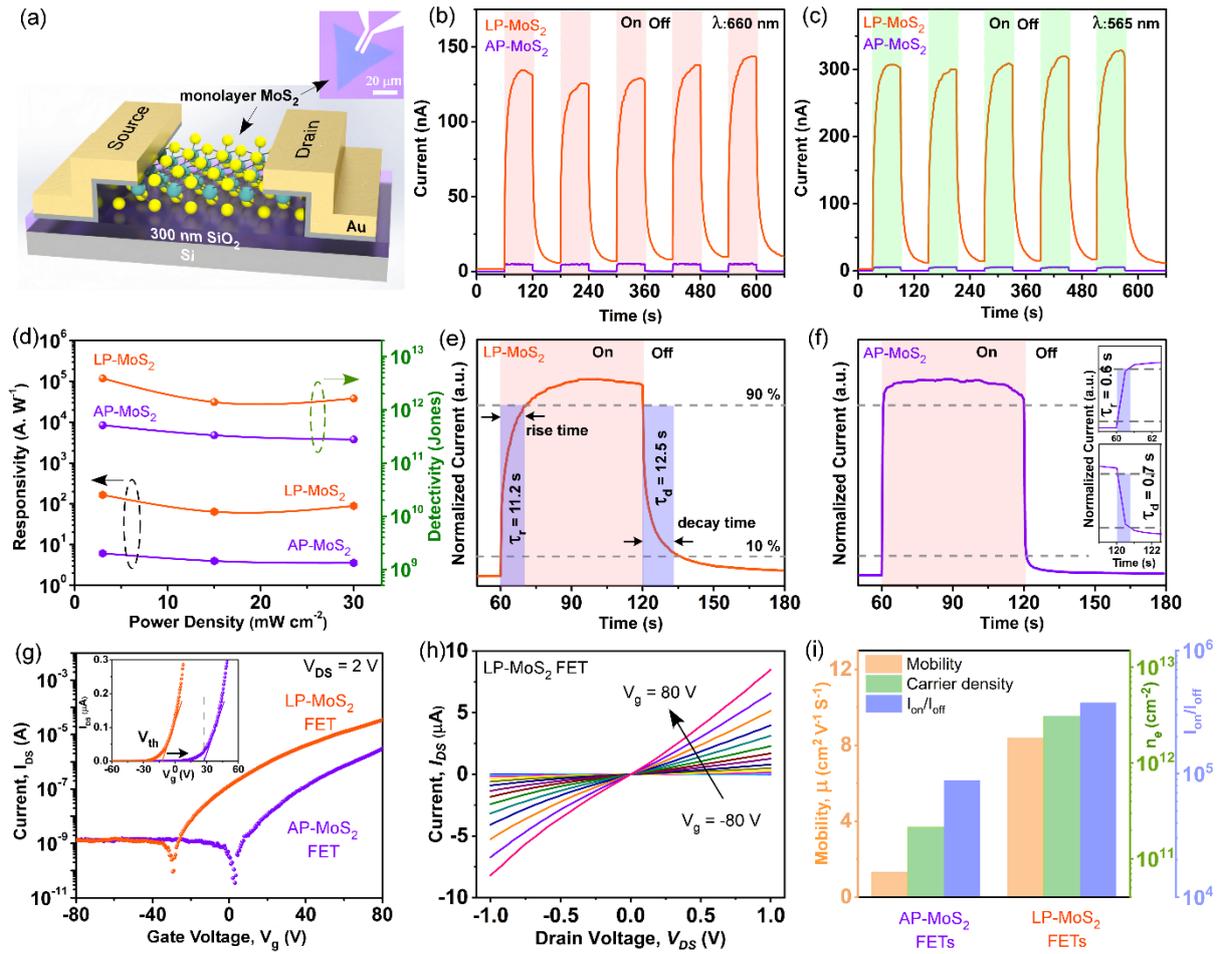

**Figure 4.** Device testing of AP-MoS$_2$ and LP-MoS$_2$ based Photodetectors and FETs. (a) Schematic of device fabricated from monolayer AP-MoS$_2$ and LP-MoS$_2$ on 300 nm SiO$_2$/Si substrate. The inset shows the optical image of a representative fabricated device. The channel length is ~ 3 μm. (b-c) Time-resolved photoresponse of AP-MoS$_2$ and LP-MoS$_2$ based devices under 660 nm and 565 nm illumination, 4 mW cm$^{-2}$ power density and at V$_{DS}$ of 2V. (d) Responsivity and detectivity of the AP-MoS$_2$ and LP-MoS$_2$ devices plot against different power density. (e-f) Plots for normalized photocurrent from AP-MoS$_2$ and LP-MoS$_2$ devices, respectively, to evaluate rise time ($\tau_r$) and decay time ($\tau_d$) of the photocurrent generated. (g) Transfer characteristics of AP-MoS$_2$ and LP-MoS$_2$ based FETs, the inset show the shift in threshold voltage (V$_T$). (h) The output characteristics (I$_{DS}$-V$_{DS}$) of LP-MoS$_2$ FETs under applied gate voltage of -80 V to +80 V showing ohmic behaviour. (i) comparison of field effect electron mobility, carrier density and I$_{on}$/I$_{off}$ for AP- MoS$_2$ and LP-MoS$_2$ FETs.

Other than photoresponsivity and detectivity, the effect of defects in MoS$_2$ is evident also from the response times i.e., rise time ($\tau_r$) and the decay time ($\tau_d$) following light exposure. Devices based on LP-MoS$_2$ show significantly longer $\tau_r$ (11.2 s) and $\tau_d$ (12.5 s), compared to AP-MoS$_2$ devices, which exhibit faster $\tau_r$ (0.6 s) and $\tau_d$ (0.7 s), respectively, as shown in Figure 3e-f. This interesting phenomenon can again be attributed to the presence of in-gap states induced by V$_s$ defects because these states act as a trap site for the photogenerated carriers leading to longer rise and decay times.[12, 17] The absence of in-gap states within AP-MoS$_2$ and the related negligible amount of traps for photogenerated carriers, leads to a faster device response. Remarkably, our simple strategy of manipulating the defects in monolayer MoS$_2$ provides a control over the photoresponse behaviour of the devices, which can be leveraged to customize optoelectronic devices. For instance, LP-MoS$_2$ devices are well tailored for emulating biological synaptic activities such as optoelectronic synaptic devices feasible for neuromorphic computation owing to the presence of persistent photoconductivity (PPC).[52, 53] Conversely, in the realm of photodetectors and sensors, where rapid response and detection are imperative for the device performance,[51] a faster response can be readily attained through AP-MoS$_2$ based photodetectors.

Furthermore, we also investigated defect engineered MoS$_2$ via this approach can be used to tune the FET properties, as presented in Figure 4g-i. The transfer characteristics of the representative devices (Figure 4g) show that the LP-MoS$_2$ FETs exhibit higher conductance in terms of on-state current ($I_{on}$) as compared to that of AP-MoS$_2$ devices. Moreover, a negative threshold voltage ($V_T$) is observed for LP-MoS$_2$ FETs (~ -12 V), whereas $V_T$ for AP-MoS$_2$ FETs shifts towards positive voltages (~ 27 V). This difference in $V_T$ between LP-MoS$_2$ and AP-MoS$_2$ devices can also be attributed to the Fermi-level shift owing to defect modulation in MoS$_2$,[43, 49] which is in alignment with our KPFM measurements, where a lower work-function value (~ 4.78 eV) is observed for LP-MoS$_2$ as compared to that of AP-MoS$_2$ (~4.95 eV). The observation of a negative $V_T$ implies that the LP-MoS$_2$ device is n-doped, while the subsequent shift of AP-MoS$_2$ devices towards positive voltage signifies the elimination of the n-doping effect attributed to the oxygen passivation of $V_s$. This observation is in accordance with the theoretical and experimental findings discussed above. In addition, the n-doping of LP-MoS$_2$ induces higher charge carrier density at a particular gate voltage,[35] which is confirmed by our calculations. An order of magnitude higher charge carrier density is observed in LP-MoS$_2$ ($n_e$ = 3.09 x 10$^{12}$ cm$^{-2}$) in comparison to that of AP-MoS$_2$ ($n_e$ = 2.16 x 10$^{11}$ cm$^{-2}$) devices shown in Figure 4i. The difference in charge carrier density between the two variants of MoS$_2$ is in accordance with the value estimated from A$_{1g}$ Raman band shift (discussed earlier).

The output current-voltage characteristics ($I_{DS}$-$V_{DS}$) of LP-MoS$_2$ FETs measured at room temperature, shown in Figure 4h, exhibit an ohmic behaviour, which is distinctly different from the non-linear Schottky characteristic observed in AP-MoS$_2$ devices, as shown in Figure S7. This indicates a relatively reduced contact resistance and a smaller Schottky barrier height (SBH) for LP-MoS$_2$ transistors with Ni/Au contact,[19] as compared to that of AP-MoS$_2$ devices. The SBH is determined by Schottky–Mott rule, but here the Fermi level pinning effect could play a dominant role,[29, 54] given that a difference in Fermi level relative to the conduction band is observed in LP-MoS$_2$ and AP-MoS$_2$. The Fermi level pinning at the highest energy state induced by defect states, closer to the conduction band minimum of MoS$_2$ can result in lower SBH in LP-MoS$_2$.[54] Consequently, the smaller SBH, lower contact resistance, and a higher charge carrier density accumulatively contribute to an enhanced $I_{on}$ and field effect mobility in LP- MoS$_2$ FETs as compared to AP-MoS$_2$ ones. Figure 4i shows that LP-CVD monolayer MoS$_2$ FETs demonstrate an average field-effect electron mobility of > 8 cm$^2$ V$^{-1}$ s$^{-1}$, which is significantly higher than AP-MoS$_2$ FETs (~ 1.3 cm$^2$ V$^{-1}$ s$^{-1}$), and is comparable to previously reported values for a typical CVD MoS$_2$ FETs.[43, 55] These findings clearly illustrate that the device performance and functionality can be enhanced and modulated by our proposed one-step CVD approach for defect engineered monolayer MoS$_2$ and other TMDs.

Finally, we have demonstrated that our proposed approach of in-situ defect engineering could be an effective strategy to tailor the activity of the basal planes of MoS$_2$ and other TMDs toward electrocatalytic hydrogen evolution reaction (HER). We employed scanning electrochemical cell microscopy (SECCM), which is a powerful technique to measure and visualize electrochemical activity with high spatiotemporal resolution.[56] This technique provides the unique capability to exclusively assess the localized HER activity on the basal plane of single-crystal 2D catalysts, as illustrated in the schematic shown in Figure 5a. This involves the voltammetric-hopping of a nanoscale pipette probe (~700 nm in diameter, herein) across the

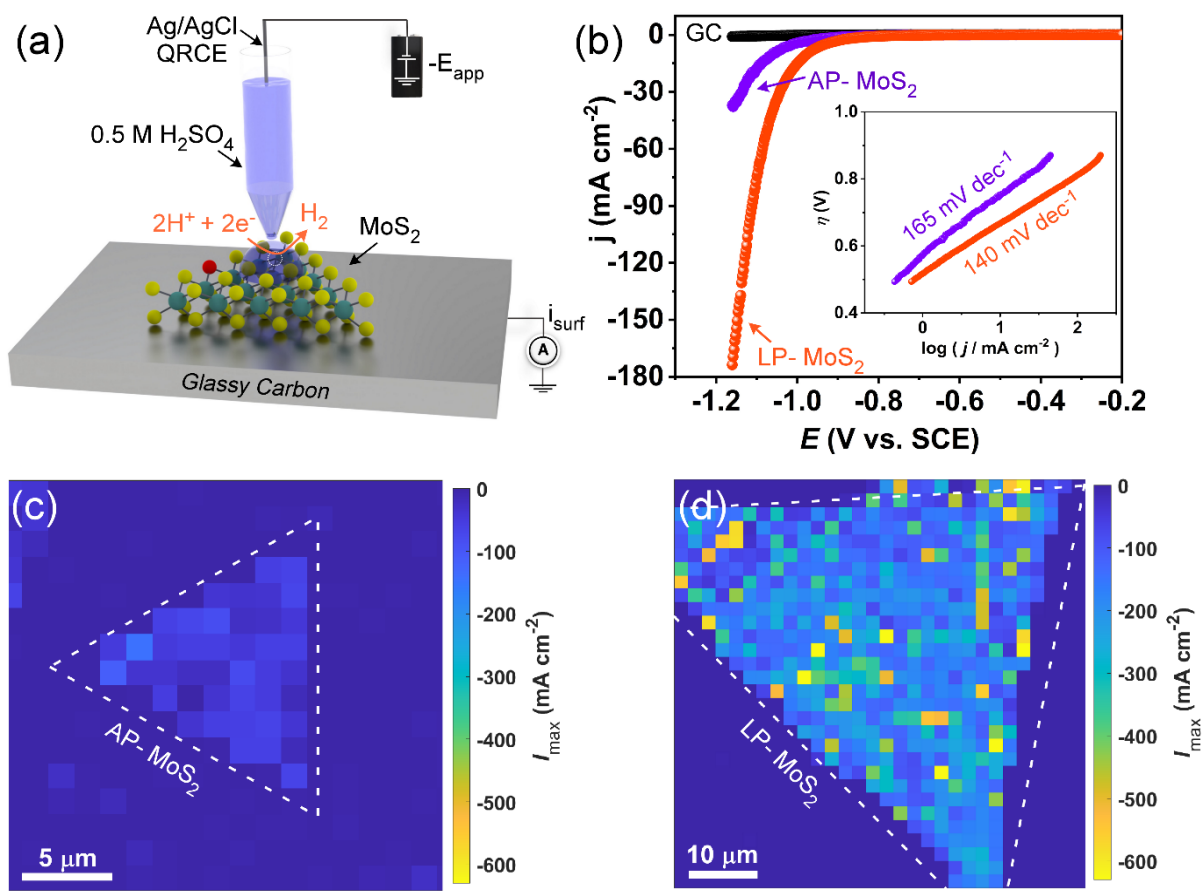

**Figure 5.** HER electrocatalytic activity of LP-MoS$_2$ and AP-MoS$_2$ (supported on GC), analysed with SECCM. (a) Schematic of voltammetric hopping mode SECCM, where $E_{app}$ is the applied potential and $i_{surf}$ is the measured surface current. (b) Representative LSVs (voltammetric scan rate = 1 V/s) obtained by averaging spatially resolved measurements taken on monolayer LP-MoS$_2$ ($N$ = 1101), AP-MoS$_2$ ($N$ = 91), and the GC support ($N$ = 166). Shown inset are Tafel plots (overpotential, $\eta$ vs. log|j|) constructed for LP-MoS$_2$ and AP-MoS$_2$, with Tafel slopes indicated. (c-d) Spatially-resolved electrochemical activity maps of (c) AP-MoS$_2$ (14 × 14 pixels, 1.5 μm hopping distance) and (d) LP-MoS$_2$ (30 × 30 pixels, 2 μm hopping distance), showing the measured current densities at $E$ = −1.18 V vs. SCE. All SECCM measurements were performed with a nanopipette probe of ~700 nm diameter, filled with 0.5 M H$_2$SO$_4$ and equipped with an Ag/AgCl QRCE.

surface of a working electrode (i.e., MoS$_2$ on glassy carbon, GC), generating an individual linear sweep voltammogram (LSV) at each measurement point. Figure 5b shows the LSVs of LP-MoS$_2$, AP-MoS$_2$ and bare GC (as reference), measured in the potential ($E$) range −0.1 to

−1.2 V vs. SCE. Direct comparison of representative LSVs (taken by averaging all pixels across the surfaces of the respective MoS$_2$ particles) indicates a clear increase in HER activity on the LP-MoS$_2$. For instance, maximum current densities measured at $E = -1.18$ V vs. SCE on the LP-MoS$_2$ (~180 mA/cm$^2$) are 5-fold greater than that of AP-MoS$_2$ (~35 mA/cm$^2$), as well as the potential required to achieve a current density of 10 mA/cm$^2$ for the HER being significantly reduced from −1.11 V vs. SCE on AP-MoS$_2$ to −1.05 V vs. SCE on LP-MoS$_2$. Tafel analysis (Figure 5b, inset) also confirms the higher activity of LP-MoS$_2$, with the Tafel slope decreasing from 170 mV/decade (AP-MoS$_2$) to 140 mV/decade (LP-MoS$_2$) with defect engineering. Note that due to the relatively high voltammetric scan rates (1 V/s, herein) and low currents measured during SECCM, the typical current density ranges used for Tafel analysis (i.e., 0.01 – 1 mA/cm$^2$, corresponding the currents of 0.03 to 3 pA, respectively, herein) are not readily accessible. Additionally, given that sulfur deficient active sites are atomic in scale, each individual spot measurement with SECCM will not only include multiple sulfur vacancies but also a large response from the relatively inactive pristine basal plane of MoS$_2$.

The combination of LSV measurements in a grid provides a spatially-resolved electrochemical 'map', through which the activity of AP-MoS$_2$ and LP-MoS$_2$, can be compared as shown in Figure 5c and d. Note that since these measurements were performed on an atomically smooth MoS$_2$ surface, all currents have been converted to current densities (i.e., intrinsic activity) herein through normalisation to the tip area of the used nanopipette probe (~4 × 10$^{-9}$ cm$^2$). These electrochemical maps clearly shows that the LP-MoS$_2$ possesses a higher overall HER activity, presenting with highly active catalytic sites distributed across the surface, compared to the relatively uniform and lower activity of AP-MoS$_2$ (Figure 5c). Looking closer at the LP-MoS$_2$ crystal (Figure 5d), the hotspots of activity throughout the crystal are represented by the yellow/orange pixels, which have maximum current densities reaching >600 mA/cm$^2$ at $E = -1.18$ V vs. SCE. These sites are thought to coincide with increased densities of V$_s$, which are known active sites for the HER on $MoS_2$.[57] This confirms the effectiveness of defect engineering for activating the usually moderately low activity of the $MoS_2$ basal plane.[58] Although exposed undercoordinated molybdenum active sites on the edges of the crystal are also expected to have elevated activities,[59] yet do not visibly appear in the scans shown. This is due to the large probe size (~700 nm in diameter) capturing the average reactivity of the atomic-scale active edge plane, relatively low activity basal plane and almost catalytically inert GC support. Nevertheless, these measurements are indeed representative of an overall increase in the reactivity or possible alternate mechanism on LP-$MoS_2$, but direct comparisons with the broader literature of macroscale (bulk) measurements,[59] should be done with caution.

**Conclusion**

In conclusion we report a novel single step approach to engineer defects in monolayer $MoS_2$ which can be further implemented across other TMDs. This method allows for selective defect passivation and enrichment within individual crystals without compromising on the material integrity. The pressure dependent growth-related defect formation/passivation is confirmed using physical, chemical and optical characterisations and theoretical calculations. We have demonstrated the effectiveness of our in-situ defect engineering strategy for customizing $MoS_2$ crystals, showcasing its applicability for specific purposes such as photodetectors, field-effect transistors (FETs), and electrochemical devices. The potential use of $V_s$ defect rich LP-$MoS_2$ for electronic, neuromorphic and catalytic devices is demonstrated through proof-of-concept devices. Whereas the use of defect passivated AP-$MoS_2$ is showcased for photoluminescence and fast photodetection applications. Hence, this study provides an effective pathway for engineering structure-property relationship of 2D layered TMDs during scalable CVD growth for integration across a wide array of applications.

**Methods:**

*CVD Synthesis of MoS$_2$*

A two-temperature zone tube furnace was used for CVD growth, in which the growth pressure was modulated by two configurations. For AP-MoS$_2$ growth, atmospheric-pressure conditions (760 torr), was maintained by allowing the downstream gas outlet to naturally vent through an exhaust. In contrast, for LP-MoS$_2$ growth, low-pressure conditions (~1 Torr) were maintained by controlling the gas outlet with a rotary vacuum pump. 300 nm SiO$_2$/Si substrates were ultrasonically cleaned by acetone, IPA and water followed by N$_2$ blow drying. An aqueous solution of ammonium molybdate tetrahydrate was prepared and was drop-casted (5μl droplet) onto the cleaned SiO$_2$/Si substrates. After that the substrates were baked at 110 °C for 5 minutes and then loaded into zone-2 of the tube furnace, under the constant flow of 500 sccm Ar for 10 minutes. For both the growth conditions, 200 mg of sulfur (S) powder was placed in zone-1, and heated to 180 °C to create S vapours, which were carried downstream by 70 sccm Ar flow to the growth zone-2, which was maintained at 750 °C during growth of 20 minutes.

*Device Fabrication:*

Two terminal lateral devices were fabricated on as grown MoS$_2$ crystals onto 300 nm SiO$_2$/Si substrate, employing standard photolithography procedures, AZ5214E photoresist (MicroChemicals GmbH) is spin coated, followed by the source/drain electrodes patterning with a maskless aligner (MLA150—Heidelberg Instruments), and developing. 100 nm Au/10 nm Ni metal contacts were deposited via electron beam deposition (PVD75—Kurt J. Lesker) with base pressure of <5 × 10−7 Torr, followed by lift-off process in acetone.

*Material Characterisation*

A Leica microscope was used to capture the optical images of the as-grown monolayer MoS$_2$ and the MoS$_2$ based devices. Thickness of the different variations of the grown MoS$_2$ crystal was determined using Bruker Dimension Icon AFM with ScanAsyst-air tip. JEOL JEM-F200 CFEG TEM with an acceleration voltage of 200 kV was utilised for physical characterisations i.e. TEM, HRTEM images and SAED patterns. XPS data was acquired for the analysis of elemental composition of the defect enriched and passivated $MoS_2$ using Kratos AXIS Supra XPS spectrometer with 1486.7 eV (Al Kα) X-ray source. The UV-vis characteristics of the different variations of $MoS_2$ was acquired using Microspectrophotometer-CRAIC. Raman and PL analysis of the different variations of $MoS_2$ was done using HORIBA LabRAM Raman spectrometer with a 532 nm laser source.

*Optoelectronic and Electronic measurements:*

Optoelectronic measurements were done using a Keysight 2912A source measurement unit on a probe station (Signatone) at ambient conditions (in air at room temperature). The devices were excited with uncollimated monochromatic LEDs (Thorlabs Inc., emitter size: 1–2.5 mm) of two different wavelengths (565 and 660nm) at a power intensity of 3 $mW/cm^2$ (calibrated with a commercial photodetector, Newport Corporation) to measure its photodetection capabilities. For the FETs characterization, back-gate device configuration was used, while transferring the as-grown LP-$MoS_2$ and AP-$MoS_2$ samples onto 285 nm $SiO_2$/Si ($p^{++}$) substrates using polymer assisted transfer method. The FET characteristics were evaluated using Keithley 4200A-SCS Parameter Analyzer on an Everbeing probe station.

*DFT calculations:*

Hybrid Density Functional Theory (H-DFT) calculations were performed using Gaussian basis set ab-initio package CRYSTAL17 with a HSE hybrid exchange-correlation functional and a double zeta basis set.[60-62] All these basis sets are given in the CRYSTAL17 basis set library. The Brillouin Zone was sampled using a 9x9x9 Monkhorst-Pack k-pt mesh for the bulk and an 9x9x1 k-pt mesh for the slab calculations. For the DFT calculations a supercell of bulk $MoS_2$ was constructed using the crystallographic information for $MoS_2$ found in the Crystallography Open Database COD ID file 9007660. Grimme D3 corrections were used to account for the dispersion contributions to the energy. After the bulk supercell was geometry optimised, a 5x5 monolayer surface slab of $MoS_2$ was cut from the bulk and structure was geometry optimised again. The monolayer structure was also doped with oxygen and sulfur vacancies were created. Total and atom projected electronic density of states plots for the (i) undoped, (ii) O-doped and (iii)S-point vacancy monolayer structures were taken calculated from the geometry optimised structures.

*Scanning Electrochemical Cell Microscopy (SECCM) Measurements*

Details on the SECCM setup, environmental control, experimental protocol and post scan data processing are available elsewhere.[58] Specific to the study carried out herein, the working electrodes were prepared by transferring AP-$MoS_2$ or LP-$MoS_2$ crystals on GC plates (10 × 10 $mm^2$, Redox.me, Sweden) that were attached to SEM pin stubs (Microscopy Solutions Pty. Ltd., Australia) using conductive silver paint (Microscopy Solutions Pyt. Ltd., Australia) to ensure a robust electrical connection. Filamented borosilicate capillary tubes (BF100-50-10, Sutter Instruments, USA, dimensions: o.d. 1.0 mm; i.d. 0.5 mm; length 100 mm) were used to produce the SECCM probes in a P-2000 laser puller (Sutter Instruments, USA). Parameters for pulling the probes are as follows; Heat – 350, Filament – 3, Velocity – 40, Delay – 220, Pull – 0 (inner diameter, 690 ± 80 nm). After pulling, these probes were filled with 0.5 M $H_2SO_4$ (Sigma-Aldrich, ACS Reagent, 95-98%), prepared using ultrapure water (Direct-Q Water Purification System, Milli-Q, USA) and equipped with an Ag/AgCl quasi-reference counter electrode (QRCE). The QRCE was prepared by the anodization of Ag wire (Goodfellow, UK: diameter, 0.125 mm; purity, 99.99%) in a saturated solution of KCl. The potential of the QRCE was calibrated against the saturated calomel electrode (SCE) (CHI150, CH Instruments, Inc., USA), against which all potentials are reported herein.

**Acknowledgements**

This work was performed in part at the Micro Nano Research Facility at RMIT University in the Victorian Node of the Australian National Fabrication Facility (ANFF). The authors acknowledge the facilities and technical support of the staff at MNRF and the RMIT Microscopy and Microanalysis Research Facility (RMMF). SPR is supported by the Australian Research Council (project number CE170100026). This work was supported by computational resources provided by the Australian Government through the National Computational Infrastructure (NCI) National Facility. SW, EDG, YL and IHA acknowledge project support via Australian Research Council Discovery Project DP220100020. SPG acknowledges support through the National Intelligence and Security Research Grant (NS210100083) and a Defence Science Institute grant. CLB acknowledges Australian Research Council (ARC) Discovery Early Career Researcher Award (DECRA, project number: DE200101076), funded by the Australian Government.